\documentclass[twocolumn,prb,showpacs,preprintnumbers,amsmath,amssymb]{revtex4}
\usepackage{textcomp}
\usepackage{wasysym}
\usepackage{amssymb}
\usepackage{graphicx}% Include figure files
\usepackage{dcolumn}% Align table columns on decimal point
\usepackage{bm}% bold math
\usepackage{verbatim}
\usepackage{color}
\begin{document}

%\preprint{APS/123-QED}

\title{Microwave-assisted transport via localized states in degenerately doped Si single electron transistors}

\author{A. Rossi\footnote[1]{Electronic mail: ar446@cam.ac.uk} and D.G. Hasko\footnotemark[2]}

\affiliation{%
\footnotemark[1]Microelectronics Research Centre, University of Cambridge, J.J. Thomson Avenue, Cambridge, CB3 0HE, U.K.\\
\\
\footnotemark[2]Centre for Advanced Photonics and Electronics, University of Cambridge, J.J. Thomson Avenue, Cambridge, CB3 0FA, U.K.\\%
}%

\date{\today}% It is always \today, today,
\begin{abstract}
Resonant microwave-assisted and DC transport are investigated in degenerately doped silicon single electron transistors. A model based on hopping via localized impurity states is developed and first used to explain both the DC temperature dependence and the AC response. In particular, the non-monotonic power dependence of the resonant current under irradiation is proved to be consistent with spatial Rabi oscillations between these localized states. 
\end{abstract}

\pacs{03.67.Lx, 71.30.+h, 71.55.Gs, 72.10.Fk, 72.15.Rn, 72.20.Ee, 73.20.Fz, 73.23.Hk}% PACS, the Physics and Astronomy
\maketitle

\section{\label{sec:intro} Introduction}

The single electron transistor is a widely used device technology for the exploitation of quantum mechanical effects (Coulomb blockade) for applications including metrology~\cite{zimm} and quantum information processing.~\cite{computing}  The first devices were fabricated using a combination of a metal (aluminium, to form the leads and the island) and an insulator (aluminium oxide, to form the tunnel barriers);~\cite{fulton,aluminium} such devices are known as metallic single electron transistors (metal SETs).  Much of the understanding of single electron effects in this type of device was obtained by measuring their electrical characteristics as a function of temperature and magnetic field.  Today, metal SETs have a well understood behaviour, exhibiting regular Coulomb oscillations (constant gate period and peak height) and a temperature dependence that is consistent with the charging energy given by the experimental Coulomb gap.\\\indent
The difficulty in fabricating these lithographically defined metal island SETs has largely limited their operation to very low temperatures (usually requiring a dilution refrigerator), so that many other material combinations have been explored to find an alternative fabrication route.  One of the most successful has been the degenerately doped silicon single electron device, fabricated using silicon-on-insulator technology.~\cite{smith-ahmed} This material system was used to demonstrate some of the earliest examples of single electron classical logic~\cite{fuji} and memory~\cite{stone} in semiconductor devices, as well as providing a route towards qubit implementation.~\cite{gorman}  This approach offers significant scalability and operating temperature advantages compared to most metal SET systems.\\\indent
In contrast with metal SETs, the electrical characteristics of degenerately doped silicon SETs usually exhibit a disordered characteristic (varying gate period and peak height). Furthermore, single electron transistor characteristics are observed without the need for a different material to form the tunnel barriers, as all of the island, the leads and the tunnel barriers are made from degenerately doped silicon.  In semiconductor materials a very effective barrier can be formed by free carrier depletion, however such a depletion process is significantly constrained by the very high doping density used to fabricate these devices.  The doping densities used for degenerately doped silicon SET fabrication are very much higher than situations where depletion is usually employed.  The minimum doping level is set by the need to be above the Mott-insulator transition ($\approx$~5\texttimes 10$^{18}$cm$^{-3}$ for P doped silicon) to avoid carrier freeze-out at low temperature and the maximum doping level is set by the solid solubility limit ($\approx$~10$^{21}$cm$^{-3}$ for P doped silicon). Coulomb blockade devices have been successfully demonstrated over the whole of this doping density range~\cite{augke,tilke} with lateral dimensions differing only by a factor of $\sim$3.  It is interesting that a change in doping density of more than two orders of magnitude makes little difference in the size or geometry of devices needed to show Coulomb blockade characteristics.\\\indent
The earliest silicon SETs were fabricated with constrictions to define the location and strength of the tunnel barriers,~\cite{augke,single} but it was soon found that these were un-necessary for the purpose of demonstrating Coulomb blockade characteristics.~\cite{alteb}  At that time, it was thought that fluctuations in the surface potential on the side walls of the silicon were the origin of the tunnel barriers.  This idea was supported by the change in characteristics observed each time the device was cycled between the room ambient and the measurement temperatures, on the principle that the surface traps would be refilled in some random way each time.  Further support for this idea comes from the observation that the temporary application of a large bias voltage to the substrate (particularly during cool-down) is effective in shifting the overall characteristics of the system, say to decrease conduction in a device where the Coulomb peaks are poorly defined.~\cite{single2}\\\indent
The disorder in the Coulomb peaks makes it difficult to extract an unambiguous charging energy.  Devices of this type can show a very wide range of charging energies, within the same transistor or within a batch of similar devices, even when fabricated together.  Furthermore, in contrast to metal SETs, their temperature dependence is not well predicted from the charging energy and standard theory.  Indeed, devices made in this laboratory have experimentally demonstrated maximum operating temperatures from a few tens of Kelvin up to nearly room temperature, despite being fabricated to the same specifications and from the same material.\\\indent
Finally, the more recent observation of high quality factor resonances during broadband microwave spectroscopy is another dramatic difference between the behaviour of degenerately doped silicon SETs and metal SETs.  Previous measurements on metal SETs showed only low quality factor resonances (quality factor $\apprle$~100) due to standing waves within the waveguide system used to couple the microwave source to the device under test.~\cite{mansch}  These standing waves cause the amplitude of the microwave signal at the device to vary with frequency and the device current is influenced by this amplitude through heating effects.  The current passed by a device under fixed source-drain and gate bias increases with temperature in a characteristic way due to the co-tunnelling effect.  This is confirmed by the experimental observation in the case of metal SETs that the device current due to the microwaves always increases and that the frequency dependence is unaffected by the bias conditions of the device under test.  This is in contrast to the experimental observations on the degenerately doped silicon SETs, where both increases and decreases in device current due to the microwaves were seen and the frequency dependence was influenced by the device bias conditions.~\cite{lisa_jap}  The quality factors of these resonances can be very high (up to around 400,000) and the peak shape varies from positive/negative gaussian to differential.  The very large number of high quality factor resonances (more than 1000 seen in the frequency range 1-20GHz) and the Rabi-like oscillations under pulsed excitation conditions,~\cite{lisa} strongly indicates a quantum mechanical origin for these effects.\\\indent
In this paper, we attempt to explain the behaviour of degenerately doped silicon SETs on the basis of localized transport resulting from significant charge trapping on the sidewalls during cool-down.  This charge trapping is influenced by the details of the device fabrication and can explain the relative independence of the required device geometry on the doping density. We will firstly describe the observed temperature dependence and disordered Coulomb blockade response at DC in terms of hopping transport via impurity states. Next, we will discuss how trapped charges can give rise to high quality factor resonances during microwave spectroscopy and what characteristics these resonances might be expected to exhibit. 
\begin{figure}[]
\includegraphics[scale=.95]{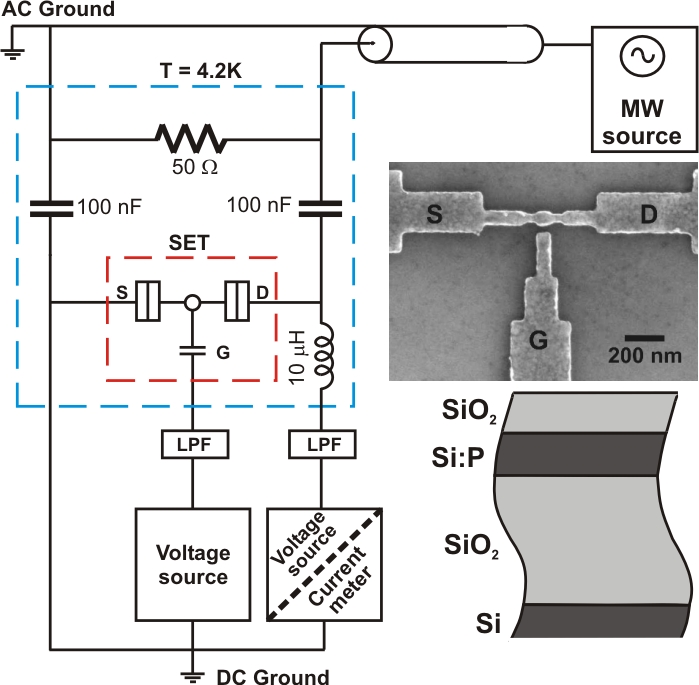}
\caption{\label{fig:circ} Schematic diagram of the radiation coupling method using separate DC and AC lines. The DC lines are low pass filtered (LPF) by MiniCircuits BLP1.9 units. Top inset: SEM image of a typical SET. Bottom inset: schematic cross section of the silicon-on-insulator structure.}
\end{figure}\indent 
\section{\label{sec:setup}Experimental details}
Devices investigated in this work were fabricated using degenerately P-doped silicon-on-insulator wafers (see bottom inset of Fig.~\ref{fig:circ}). Devices were fabricated using electron beam lithography and a sacrificial metal mask to transfer the pattern by reactive ion etching. Patterned substrates were thermally oxidized at 1000\textcelsius~to reduce the surface trap state density, which in turn reduces the incidence of random telegraph signals. A micrograph of a typical measured SET is depicted in the top inset of Fig.~\ref{fig:circ}.\\
\indent Measurements were carried out in a specially constructed probe that could be directly immersed in a liquid cryogen, so that the device under test and a substantial part of the connecting leads, were all efficiently cooled.  All measurements discussed here were performed in liquid helium, at a temperature of 4.2K. DC measurements were made using Keithley 236 source-measure units (SMU) controlled by a LabView programme via the GPIB bus. The DC measurement connecting leads were directly coupled to the source, drain and gate terminals of the device and individually low pass filtered by MiniCircuits BLP1.9 inline units at room temperature with cut-off frequency of 1.8 MHz.  This filtering is essential to prevent electromagnetic noise (due to external sources) from propagating to the device under test, which would otherwise be subject to strong electron heating. Microwave signals were provided by an Agilent E8257D source capable of delivering up to +15dBm via a flexible sma-terminated cable to the probe.\\\indent AC signals can be applied to the source-drain leads via a biasing circuit; a schematic of the coupling circuit is shown in Fig.~\ref{fig:circ}. A 50~$\Omega$ metal film resistor is provided to impedance match the waveguide and microwave source, so that standing waves are inhibited.  100 nF ceramic capacitors are used to decouple any DC voltage on the waveguide from reaching the device.  At 100 MHz (the lowest frequency used here) the impedance of each capacitor is $<1~\Omega$ and  decreases with increasing frequency   so that the voltage drop across the 50$\Omega$  load resistor appears as a voltage difference between the source and the drain sample connecting leads, with negligible loss.  The 10~$\mu$H inductor (impedance $>6~$k$\Omega$  at 100~MHz) prevents the AC voltage from reaching the SMU, where it could interfere with the DC current measurement.\\\indent
Investigations have been previously carried out where the radiation was indirectly coupled to the sample by broadcasting it from an open-ended wave-guide.~\cite{lisa_jap} This indirect coupling method resulted in a global time dependent potential for the whole device, whereas the more efficient coupling technique presented here results in a time dependent potential difference between device electrodes (mainly source and drain). 
\begin{figure}[b]
\includegraphics[scale=0.51]{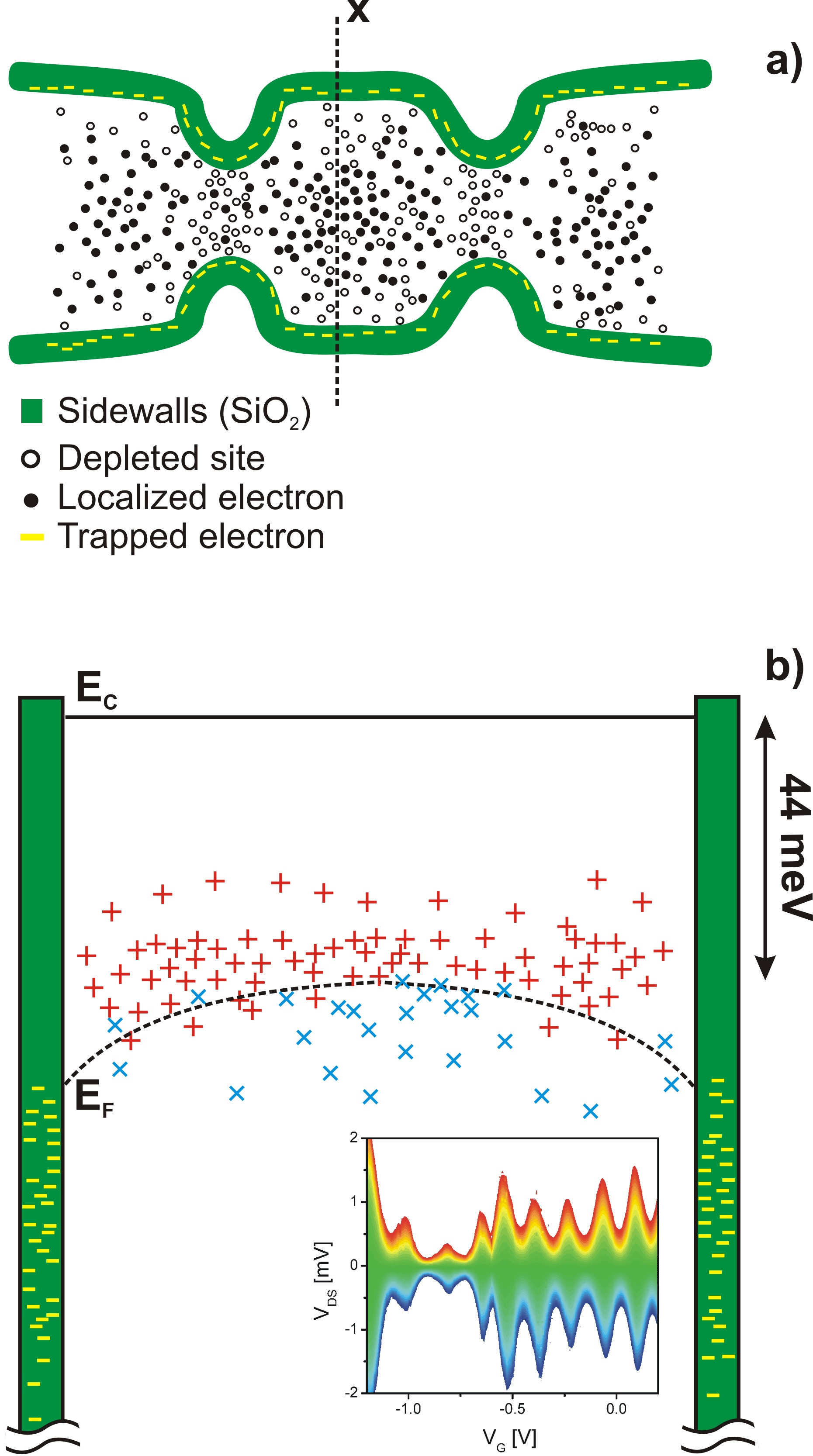}
\caption{\label{fig:energies}Depletion effect due to sidewalls trapping. a) Cartoon showing a top-view of the nano-structure. Depleted regions are mostly formed in the vicinity of the side edges and where the channel is narrower. b) Cross sectional electrostatic potential at the generic position marked by the dotted line in (a). Fermi energy is bent because of charge accumulation at the sidewalls. Ionised donors and neutral sites with bound electrons are represented by plusses and crosses, respectively. Inset: a device stability plot: drain-source current as a function of drain-source voltage, V$_{DS}$, and side gate voltage, V$_G$. Blue-to-red scale corresponds to current varying from -60 pA to +60 pA.}
\end{figure}

\section{DC transport}
\begin{figure}[]
\includegraphics[scale=0.420]{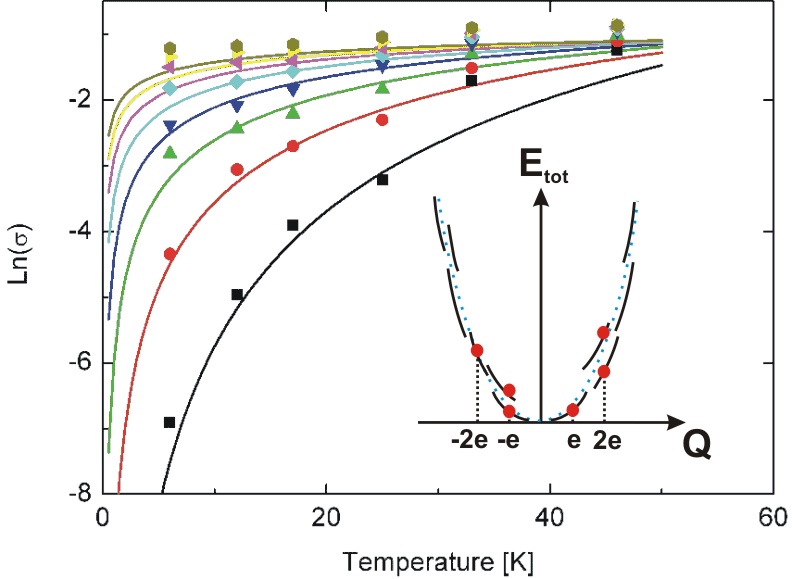}
\caption[]{\label{fig:DC}  Natural log of a device differential conductance at different temperature and electric field. Measurements were taken at applied source-drain voltages of  0, 0.02, 0.04, 0.06, 0.08, 0.1, 0.12 and 0.14V from bottom to top.  Solid lines are the results from a fit using  eq.~(\ref{eq:temp_fit}). Inset: total energy on the island as a function of the stored charge for a classical metal Coulomb Blockade device (dotted line) and a highly doped silicon device (solid line). The red dots indicate positions corresponding to integer charge. Of note is that the solid parabola is segmented and displaced due to the additional contributions from the binding and interaction energies.  Some of the integer charge positions have multiple solutions corresponding to different charge configurations.}
\end{figure}

A typical stability plot, measured at a temperature of 4K, is shown in the inset of Fig.~\ref{fig:energies}(b) showing aperiodic Coulomb oscillations and unequal peak heights.  Clearly this device must meet the two requirements for the observation of Coulomb blockade; the thermal energy must be smaller than the charging energy and the tunnel barrier impedance must be large compared to the resistance quantum.  The second condition requires that the tunnel barrier impedance should be at least 100k$\Omega$  and the first condition requires that the tunnel barrier does not extend by more than one or two squares along the device axis (otherwise the capacitance of the island is too large).  However, a doping density of $\approx$10$^{19}$~cm$^{-3}$ gives a sheet resistance of  about 1600$\Omega/\square$ at room temperature, dropping by a factor of about 1.6 at 4K,\cite{smith-ahmed} giving a \textquotedblleft tunnel barrier\textquotedblright resistance of only a few k$\Omega$  from geometrical considerations alone.  Clearly, the carrier transport in the tunnel barrier region must be influenced by some other mechanism that greatly increased the resistance.  To explain the magnitude of the current observed at a Coulomb peak, the tunnel barrier resistance must be nearly two orders of magnitude larger than the value expected from the geometry and from the doping density.\\\indent
It was previously suggested that the high tunnel barrier resistance is a result of the depletion of free carriers by charge trapped at the sidewalls of the silicon.~\cite{arxiv}  Of course, a high density of traps at the sidewalls is expected due to the non-(100) orientation of the surface.~\cite{vitk}  
In general, the density of traps due to unsatisfied dangling bonds at the silicon surface is dependent on the crystalline orientation of the sidewalls. Oxidation is routinely used to passivate such dangling bonds, but little difference in device behavior was seen for the earliest devices made without oxidation and those with oxidation. Oxidation is effective in reducing random telegraph noise, but has little effect on device geometry necessary for the observation of Coulomb Blockade.  Moreover, the depletion depth due to these traps would be fixed (and estimated to be much too small to explain the tunnel barrier size) for a given doping density.  So that an increase in the doping density would require a corresponding decrease in the physical size of the tunnel barrier, contrary to experimental observation. Another mechanism for trap formation arises from the fate of dopant atoms in the silicon etched away to form the sidewalls.  Reactive ion etching is used to form the island, tunnel barriers and leads by removing the unwanted silicon.  This process proceeds by transporting gas reactants to the silicon surface, where a chemical reaction takes place to form volatile products, which are removed by the vacuum pumping system.  Typical silicon dopants, such as phosphorus, do not form volatile compounds with the CF$_4$ and SiCl$_4$ reactive etch gases at the usual etch temperatures.  The phosphorus atoms remain on the surface after etching and become incorporated into the oxide during passivation.  These atoms are no longer able to act as dopants, but do act as traps.  The density of these traps directly depends on the doping density of the silicon so that a change in this doping density has no direct effect on the depletion depth in the silicon.  This mechanism explains why all tunnel barriers appear similar for a wide range of doping density.\\\indent
Transport in highly doped silicon SETs is strongly localized in and around the SET island and tunnel barriers, due to depletion resulting from trapped charge on the sidewalls.  At high temperatures, these traps will be mostly empty due to thermal excitation; so that the resistance is little changed from that expected on the basis of the bulk material properties.  But, as the temperature is reduced these empty, neutral traps are able to capture electrons that tunnel in from the nearby silicon, which results in a significant negative charge accumulation at the sidewall.  To maintain overall charge neutrality, this negative charge causes depletion of free carriers from the nearby silicon; these being the most likely to be captured in the traps anyway.  This depletion increases the size of the tunnel barrier; but in practice, would only affect the resistance of narrow sections, such as the SET island and tunnel barriers (Fig.~\ref{fig:energies}). Such depletion moves the Fermi energy down into the impurity band, despite the high doping density, so that the conduction mechanism changes from metallic to a hopping behaviour.\\\indent
Such conduction behaviour normally arises only in lightly doped semiconductors at low temperatures when the previously free carriers are recaptured by the ionised dopants and the overlap between carrier wavefunctions on adjacent impurity sites becomes weak.  But such characteristics can also be observed in systems where the carrier density is usually high but is subject to depletion due to a gate voltage or by trapped charge. In that case, the conductivity decreases exponentially as the energy or spatial separation between the localized sites increases.  We can apply the analysis of Efros and Shklovskii~\cite{ES_book} to these situations.
\\\indent
Fig.~\ref{fig:DC} shows the electrical differential conductance measured as a function of temperature and source-drain voltage for a highly doped silicon nano-wire device, under conditions where the Coulomb Blockade effects will be mostly suppressed by the applied potential difference at drain-source and/or gate. Measurements have been performed at six different temperatures for each applied electric field by means of standard lock-in techniques. The data points shown are the average values for the differential conductance from a number of equivalent experimental conditions obtained by sweeping $V_{DS}$ in both forward and reverse directions. The observed behaviour is typical of a wide range of device types, including those with lithographically defined tunnel barriers. It is clear that both temperature and electric field have a strong influence on the transport properties.  This current dependence is often attributed to co-tunnelling in Coulomb Blockade devices, where the tunnel barrier is formed by a material with a large energy gap.  An analytic form for the inelastic co-tunnelling contribution to the sample conductivity of an SET measured at low temperature T and applied source-drain voltage V$_{DS}$ has been given by Averin and Nazarov~\cite{averin} as
\begin{equation}
\sigma_{cot}=A(BT^2+CV_{DS}^2)
\end{equation}
where A, B and C are constants that depend on the nature of the tunnel barriers and fundamental constants. However, it is clear that such a relation does not satisfactorily explain the dependencies shown, where the largest changes occur at the lower voltages and temperatures.
\\\indent
As discussed earlier, the electron density close to the sidewalls is expected to be significantly reduced by depletion resulting from electrons trapped in the grown oxide, so that it is appropriate to consider the temperature and electric field dependence of the hopping conductivity in low-doped semiconductors to explain the characteristics in Fig.~\ref{fig:DC}.  The conductivity of a disordered system in the small polaron hopping regime has been discussed by Triberis.~\cite{triberis} This work concludes that Mott's law gives the main temperature dependence and that the electric field dependence is given by a function of the square of the source-drain voltage.  Using these relations, the following function has been fitted to the data in Fig.~\ref{fig:DC}  
\begin{equation}
\label{eq:temp_fit}
ln[\sigma(T,V_{DS})]=A+BT^{-1/4}
\end{equation}
where $A$ and $B$ are treated as different fitting constants for each value of $V_{DS}$. The functions that model the field dependence are then fitted as follows
\begin{equation}
A=C+D(E+V_{DS})^{-2}
\end{equation}
\begin{equation}
B=F(E+V_{DS})^{-2}
\end{equation}
being $C=-1.06$\textpm$0.08$, $D=0.024$\textpm$0.001$, $F=-0.067$\textpm$0.001$ and $E=0.054$ fitting parameters ($E$ is used as a fixed parameter, therefore it does not come with an error). The fit for low temperatures and small source-drain voltages is very satisfactory; at higher voltages and temperatures the hopping model becomes progressively less applicable.  This agreement between the experimental data and the hopping model suggests that the observed temperature dependence is mostly determined by changes in the transparency of the tunnel barriers rather than by co-tunnelling.  
\\\indent Very large charging energies have been estimated in the past from electrical measurements at low temperatures suggesting that operation at high temperature should be possible.~\cite{augke}  However, these estimates often disagree with the observed maximum operating temperature.  In these cases, it is the degradation of the tunnel barrier resistance that is the cause of the reduction in maximum operating temperature, rather than the onset of significant co-tunnelling.  Longer constrictions seem to give higher maximum operating temperatures than shorter or narrower ones.  This effect may be explained by the increase in the number of hops taken to pass through the tunnel barrier.  An increase in the number of hops will reduce the energy barrier per hop and so reduce the effect of an increase in temperature or electric field.
\\\indent
In a metallic SET, the electrical characteristics are determined by the island capacitance and the tunnel barrier transmission.  The charging energy for an extra electron to occupy the island has only one contribution, that due to the island capacitance.  In a degenerately doped silicon SET, hopping transport can modify this behaviour, since the energy for an electron to occupy a localized site in the island now has three contributions.  The first is the conventional Coulomb charging energy due to an amount of charge Q on an island capacitance C (we will assume a fixed capacitance), where the localized site is situated.  The second contribution is due to the binding energy of the site, which will depend on the nature of the localisation. This contribution decreases the system energy by an amount $E_{bi}$ for each electron. The binding energy for an isolated phosphorus atom in a silicon lattice is given by the Bohr model as $\approx$~44meV.  However, in degenerately doped material the donor spacing is sufficiently close that wavefunction overlap strongly influences this result.  Disorder in the donor lattice position leads to a deviation from the average donor separation, $r$, that varies with $r^3$; so that while most donors are separated by a spacing that is close to the average value, a small fraction have significantly smaller spacings.  At these locations the binding energy for one electron is increased and for the second is decreased.  The effects of the disorder in the donor location and the local influence of the trapped charge causes the binding energy to be different at each localized site (Fig.~\ref{fig:energies}).  The third contribution is due to the electron interactions between the occupied localized sites.  If the spacing between an occupied site pair is $r$, then this increases the energy by an amount $E_i$ which depends on the electron charge $e$, on the permittivity of the channel material $\epsilon$ and on $r$ in the following way,
\begin{equation}
\label{eq:interaction}
E_i=\frac{e^2}{4\pi\epsilon r}
\end{equation}
The total energy of the system E$_{tot}$, then depends on these three contributions
\begin{equation}
E_{tot}=\displaystyle\sum_{i=1}^\frac{N!}{2!(N-2)!} E_i(r)-
\displaystyle\sum_{i=1}^{N} E_{bi}(r)+
\frac{Ne^2}{2C}
\end{equation}
where $N$ is the number of electrons localized at donor sites within the island.
For the purposes of assessing the Coulomb blockade charging energy, we are only interested in the change in the total energy  $E_{tot}$ for the addition or subtraction of one electron.
\begin{equation}
\Delta E_{tot}=NE_i(r)-E_{bi}+
\frac{e^2}{2C}
\end{equation}
So that the classical Coulomb Blockade energy is modified by two additional terms.  The magnitude of the first of these additional terms (the interaction energy) will be $N$ times about 3~meV (assuming a separation of about half the typical dot diameter of $\approx$40~nm) and the second (binding energy) term will be of the order of 44~meV from the previous discussion.  Clearly, with about 15 electrons on the island these two additional terms are expected to cancel and the classical Coulomb Blockade charging energy depends only on the island capacitance.  However, as considerable variation from these average approximations are expected, then the actual charging energy will contain a correction term to the classical result, which is due to the difference between the binding energy and interaction terms.  This difference will vary as the electron number varies and for a given electron number will also vary with configuration i.e. if a different set of donors is occupied with the same number of electrons then the energy difference is also changed.  In addition, the estimate for the interaction term makes no allowance for screening, which will greatly reduce the influence of charges at the larger separations. Indeed, the screened Coulomb interaction energy will decay rapidly with distance; we describe it as follows:
\begin{equation}
E_{scr}=E_ie^{-k_sr}
\end{equation}    
where $1/k_s$ is the Thomas-Fermi screening length. We estimate $1/k_s\approx$114nm for our system; this may increase the number of electrons on the island to achieve a balance between the binding energy and interaction energy terms to the order of 100. The effect of these additional terms is illustrated schematically in the inset of Fig.~\ref{fig:DC}, where the parabolic dependence of energy on stored charge for the classical metallic Coulomb Blockade device (blue dotted line) is changed to a discontinuous, and sometimes multi-valued, approximate parabola in the highly doped silicon single electron device (black solid line). Clearly the latter case will result in aperiodic Coulomb oscillations.  Moreover, the multiple values corresponding to particular electron numbers result in the position of a Coulomb peak depending on which electron configuration is adopted (the energy barrier between configurations may prevent simultaneous occupation).  In this case, the Coulomb peak pattern can depend on the gate voltage history prior to the observation.
\section{Transport under microwave irradiation}
\begin{figure}[]
\includegraphics[scale=.8]{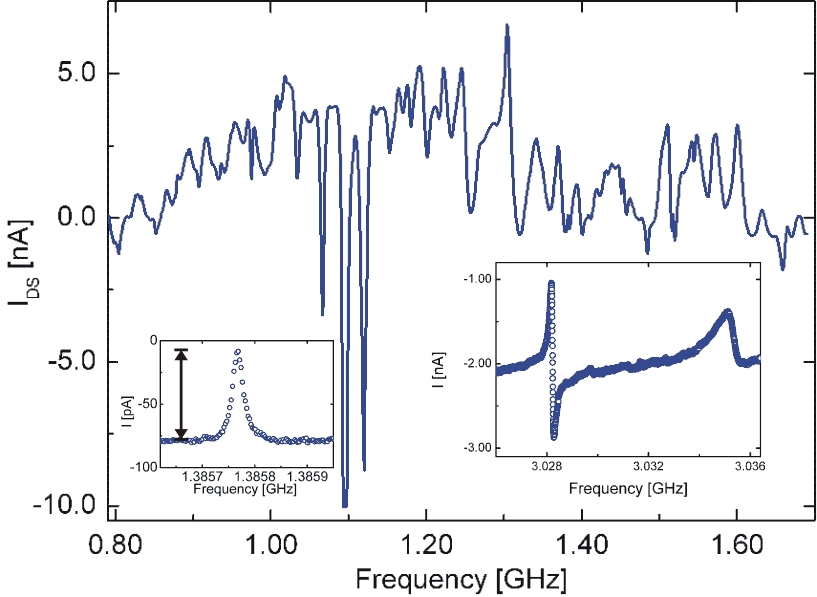}
\caption{\label{fig:res} SET response to AC excitation. The large amplitude peaks have typically low Q value (few tens). Bias voltages: V$_{DS}$=+0.7~mV, V$_G$=-2.77~V; radiation power: P=+15~dBm. Left inset: a high Q-value peak on an expanded frequency scale, with amplitude indicated by the double headed arrow. Q$\approx$10$^5$. Right inset: resonances of different line-shape and medium Q closely spaced in frequency. Different data-set with respect to the main plot. Bias voltages: V$_{DS}$=-1~mV, V$_G$=-1~V; P=~0~dBm.}
\end{figure}

\begin{figure}[b]
\includegraphics[scale=0.45]{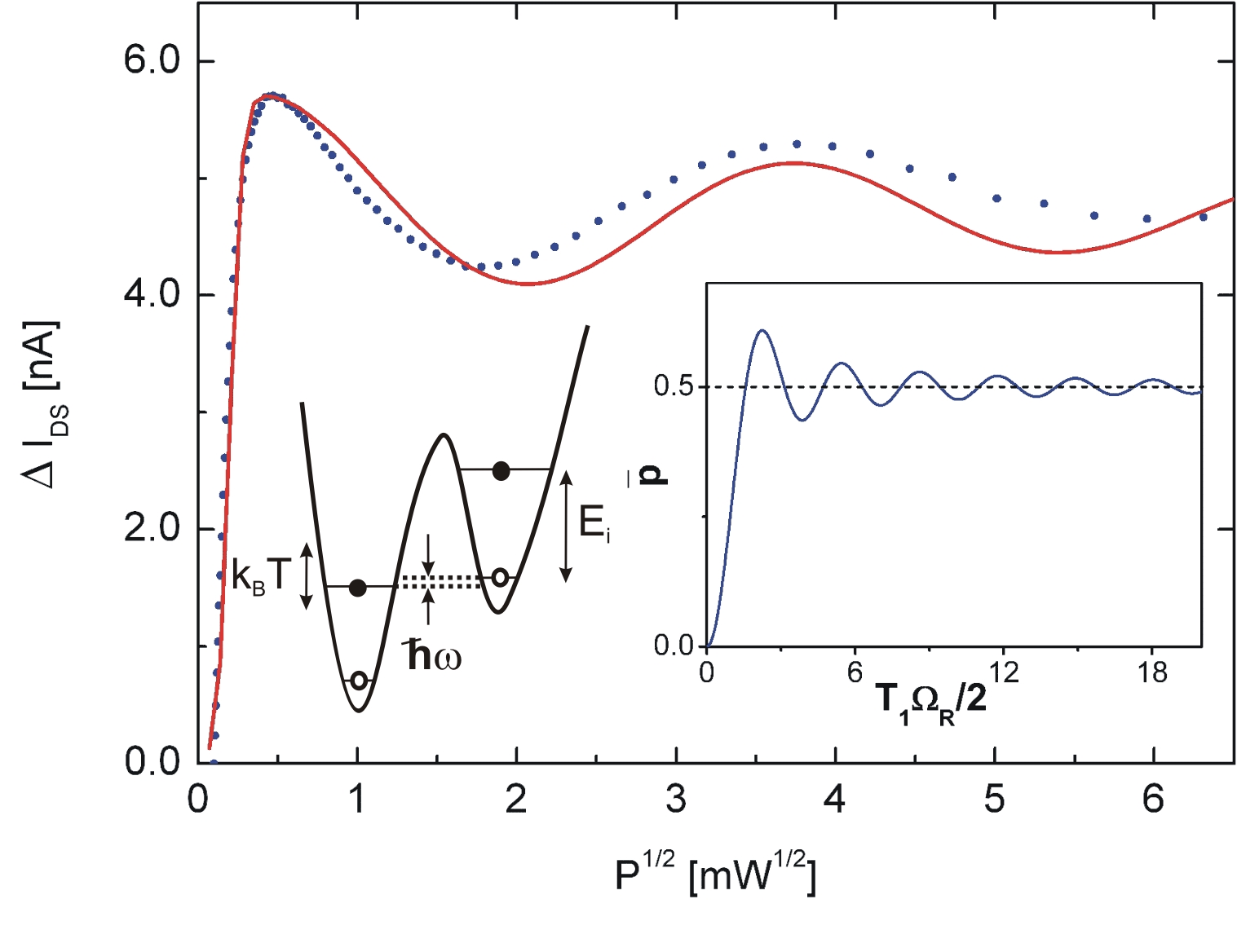}
\caption{\label{fig:power} Dotted line: experimental current dependence at resonance (Frequency=1.9200 GHz) for varying radiation power. $\Delta I_{DS}$ is the excess current with respect to DC condition. Bias conditions: V$_G$=-2.0 V, V$_{DS}$=-1.30 mV. Solid line: calculated current using eq.(\ref{eq:I_fin}) with $I_e$=9.65$\times$10$^{-9}$A, $c$=7.33$\times$10$^{12}$~$\frac{rad^2/s^2}{mW}$, $\delta$=4.96$\times$10$^6$ rad/s, $k$=4.98, $\alpha$=104.04, $T_1$=10$^{-6}$s. Left inset: schematic energy level diagram showing two donor locations and the occupied (filled circle) and unoccupied (empty circle) energy levels. Right inset: average probability of occupation of an excited state under resonant excitation against the ratio between the relaxation time and the Rabi period (eq.(\ref{eq:p_ave})).}
\end{figure}

In a typical experiment, the SET is biased with a fixed source-drain voltage and held at a fixed gate voltage, while the source-drain current is measured as the microwave frequency is swept.  This must be done with the SET set up as an impedance matched load (or with a matching impedance very close to the SET) or with the SET in a low quality factor (Q) cavity, which itself is matched to the waveguide impedance.  In the latter case, the coupling efficiency is very greatly reduced and this approach has been used in this work.  A typical result is shown in Fig.~\ref{fig:res}, where the following features may be identified; large amplitude, low Q fluctuations (main plot), medium amplitude, mid Q fluctuations (right inset), and low amplitude, high Q resonances (left inset).  
The high Q features have been extensively studied by Cresswell and coworkers, who have demonstrated that the resonant frequency is affected by both the gate and the source-drain voltages, although the effect can be very small.  Resonant features have previously been observed in two-dimensional electron gas systems (silicon MOSFETs and AlGaAs heterostructures), where this behaviour has been attributed to single particle excitation within the Fermi sea.~\cite{fujisawa} In this previous work, a magnetic field is used to determine the energy difference between two levels, giving rise to a single resonance, resulting in only small numbers of resonances due to the requirements for a separately identifiable spin active system for each resonance.  By contrast, the degenerately doped silicon SET exhibits large numbers of resonances without the need to apply a magnetic field.  The large number of resonances (many more than the expected number of free electrons in the island) suggests that their origin involves single particle excitation, with each able to give rise to a number of resonances.  Furthermore, the photon energy corresponds to a small fraction of the thermal energy at the measurement temperature of 4K, so that the populations in the two levels would be expected to be very nearly equal (of the order 0.49 and 0.51 for the upper and lower energy levels).  
\\\indent
The AC conduction behaviour of the hopping model, described earlier to explain the DC characteristics of degenerately doped silicon SETs, has been considered previously.~\cite{migliuolo,ritz}  All of these models consider statistically averaged behaviour over a large number of hops in order to predict a value for the conductance.  However, in these SETs individual hops may contribute a significant change in the conductivity due to the small size of the device.  The basic underlying mechanism is the change in hopping rate between two localized sites, when exposed to electromagnetic radiation that is resonant with the energy level difference between the sites.  This changes the occupation levels at the two sites away from the values corresponding to thermal equilibrium.  In an SET with localized sites, a change in just two sites may well give rise to a measurable change in the conductance, due to the very high sensitivity of a tunnel barrier transmission to wavefunction overlap.  Any change in occupancy of localized sites close to the edge of a tunnel barrier will have an exponential effect on the transmission.  If these changes in occupancy occur at sites further away, then the tunnel barrier is affected more indirectly through the polarisation of intermediate sites, which reduces the overall change in transmission.\\\indent
Consider two sites separated by a distance that is significantly larger than the wavefunction overlap, and by a tunnel barrier that reduces the thermally driven tunnelling rate between the two sites to a very low level, see left inset of Fig.~\ref{fig:power}. Under these circumstances, electromagnetic radiation that is resonant with the separation between these energy levels will cause the electron in the occupied level to tunnel to and from the unoccupied level, at a rate (the Rabi rate) determined by the intensity of the radiation.  If the intensity is very high, then the electron will perform these spatial Rabi oscillations, as discussed by Stafford and Wingreen.~\cite{staff}  Such oscillations can proceed coherently until there is an interaction with the environment that results in an exchange of energy and/or causes the electron to tunnel to a different site; such interactions determine the lifetime of these oscillations.  
\\\indent
The changes in site occupancy resulting from these spatial Rabi oscillations causes a change in the SET current by modifying the transmission properties of the tunnel barriers.  This modification is a result of the change in electrostatic potential resulting from the electron displacement in the oscillation.  The magnitude of this change depends on the separation between the sites, the distances and directions to the tunnel barriers and the Rabi rate.  The electron displacement due to the Rabi oscillation gives rise to a time dependent polarisation, which gives rise to the change in potential at the tunnel barrier.  However, this electric field is screened by the polarisability of any intermediate structures, such as other tunnelling charges.  If the spatial Rabi oscillation is slow, then screening can be very efficient and the affect on the tunnel barrier transmission is correspondingly small.  As the Rabi rate is increased, the screening is progressively overcome and the tunnel barrier transmission is modified.\\\indent
A single electron undergoing a spatial Rabi oscillation will cause a monodirectional resonance (left inset of Fig.~\ref{fig:res}). However, there is also a natural tendency for such systems to couple. The amplitude response of a system of coupled oscillators is well known~\cite{french} and results in a resonance whenever the driving field corresponds to a mode frequency. The individual motions of the electrons participating in the coupled mode is more complicated than the sum behavior. In particular, the relative amplitude for each coupled electron differs on either side of a resonant frequency. In cases where electrons acting independently would give rise to monodirectional resonances in opposite directions, the coupled motion would give rise to a differential behavior, as observed in Fig.~\ref{fig:res} (right inset).  
\\\indent The width of each resonance directly reflects the lifetime of the oscillation, which can vary over a very wide range.  Rabi oscillations are usually limited to a maximum rate that is very much slower than the resonant frequency of the excitation (say by a factor of 100). The lowest frequency at which high quality factor resonances have been observed in these experiments is of the order 1GHz, so that the Rabi rate probably does not exceed a few tens of MHz. In addition, if the Rabi rate is slower than the rate at which the electron interacts with the environment, then complete oscillations cannot be made and there is no resulting polarisation to change the SET current.\\\indent
As briefly mentioned above, the energy difference between levels undergoing spatial Rabi oscillations, given by the resonant frequency, is much smaller than $k_BT$ at 4K, so that the Boltzmann function predicts a very small difference between the occupancies for these levels at thermal equilibrium. The spatial Rabi oscillations, due to continuous wave  radiation, can only change the occupancy levels towards the limiting case of 50\% (right inset of Fig.~\ref{fig:power}), so if these levels were already close to this figure there would be little change in polarisation between the localized sites and so little change in device current.  However, it is important to point out that for low dimensional structures at cryogenic temperature the electron-phonon interactions are strongly suppressed.\cite{pescini,bourg} In particular, Tilke and coworkers~\cite{tilke2} have found that for P-doped silicon nano-wires the electron-phonon relaxation time can be as high as few $\mu$s. Single-shot measurements~\cite{thierry_shot} performed on the same device system investigated here have shown that electron relaxation time due to microwave excitation has a comparable timescale. Therefore, we can assume that the thermalisation processes in the dot occur on a longer timescale than the photon-induced charge transfer does. This accounts for the observation of radiation effects at energy much lower than $k_BT$.
\\\indent Finally, we turn to explain the effect of a variable radiation power on the resonant current. The Rabi rate, $\Omega_R$, increases as the square root of the  microwave power under resonant excitation conditions. The magnitude of the device current difference is thus expected to change correspondingly. As pointed out earlier, the Rabi rate must exceed the rate of interaction with environment for any net charge displacement, so that in practice there is a threshold in microwave power that must be exceeded for any resonance to appear. The threshold condition occurs when $T_1\approx 1/\Omega_R$ where $T_1$ is the relaxation time for the resonance. The occupancy of the normally unoccupied site increases with microwave power, saturating at 0.5 for very high powers. For intermediate power levels, the occupancy oscillates due to beating effects between the Rabi rate and $T_1$ (see right inset of Fig.~\ref{fig:power}). The average occupation probability of the excited state is given by
\begin{equation}
\label{eq:p_ave}
\overline{p}=\frac{1}{T_1}\int^{T_1}_0 p_e(t)\,dt 
\end{equation}
where $p_e(t)=sin^2(\Omega_R t/2)$ is the instantaneous occupancy. In order to evaluate the current dependence on radiation power (P), we assume that each of the two states corresponds to currents $I_g$ and $I_e$. The instantaneous current is given by the weighted sum of the currents,
\begin{equation}
\label{eq:I_inst}
I(t)=I_g(1-p_e(t))+I_ep_e(t) 
\end{equation}
and the average current is
\begin{equation}
\overline{I}=\frac{1}{T_1}\int^{T_1}_0 I(t)\,dt 
\end{equation}
Since we are mainly interested in the current variation, we can set $I_g$=0 which would result in 
 \begin{equation}
\Delta\overline{I}=I_e\overline{p}=I_e\frac{T_1\Omega_R/2-cos(T_1\Omega_R/2)sin(T_1\Omega_R/2)}{T_1\Omega_R} 
\end{equation}
In order to model the rapid suppression of any excitation in case $T_1\ll 1/\Omega_R$, we introduce an exponential prefactor as follows
 \begin{eqnarray}
\label{eq:I_fin}
\Delta\overline{I}= 
\frac{1}{1+e^{-\alpha(T_1\Omega_R-k)}} \\
\cdotp I_e\frac{T_1\Omega_R/2-cos(T_1\Omega_R/2)sin(T_1\Omega_R/2)}{T_1\Omega_R} \nonumber
\end{eqnarray}
with $\alpha$, $k$ fitting constants. As stated earlier, the Rabi rate is proportional to the square root of the radiation power and can be expressed as
 \begin{equation}
\Omega_R=\sqrt{cP+\delta^2} 
\end{equation}
being $c$ the constant of proportionality and $\delta$ the detuning factor. Fig.~\ref{fig:power} shows a comparison of the modelled excess current as a function of the radiation power with experimental data taken from a resonance at 1.9200 GHz and Q~$\approx$~2000. As the power is swept, a non-monotonic behavior for the resonant current is observed for the vast majority of the radiation-induced peaks and the model successfully replicates its main features. However, at the lowest (highest) power levels the model response appears to be slightly slower (faster) than the real data response. This may be because the model does not include the effects of charge screening and rectification which may play a role at low and high power regimes, respectively.
%\section{Application to QIP}
%The effects investigated here can be exploited to achieve quantum computation. We refer to the guidelines given by DiVincenzo~\cite{vince} who identifies five requirements to be met by a system suited for quantum computing: identification of qubit, state initialisation, low decoherence, state manipulation and reliable quantum measurements. We propose that each resonance can be considered as a separate qubit, characterised by the resonant feature shape, frequency and width, due to the displacement of an electron between a pair of donor states within the SET. The quantum system initial state is defined by the set(s) of occupied donor sites that minimises the free energy in the system. Spectroscopic characterisation indicates that a large number of potential qubits is available within a single device. Manipulation and gate operation can be based on well established NMR protocols. Readout is limited to single bits by addressing in the frequency domain and observing the current change as a result of the application of a $\pi$ pulse. The direction of the current change in response to this pulse indicates the state of the bit. As other bits will be also perturbed, no further readout is possible without re-running the computation first.
\section{\label{sec:end}Conclusion}
The very high doping density does not ensure metallic behavior in the tunnel barrier and the island of a silicon SET. Increasing the doping density cannot overcome the sidewalls depletion causing the non-metallic behavior due to a consequential increase in sidewall trap sites. Transport through the tunnel barriers and island is dominated by hopping through a limited number of sites. Rearrangement of a fixed number of electrons between a larger number of sites is seen to lead to aperiodic gate oscillations and varying peak heights. Individual transfers can be promoted by resonant microwave excitation, resulting in a spatial Rabi oscillation. Such transfers are detected through their effect on the DC transport. The equilibrium population and lifetime of these states is strongly influenced by the energetics of localized electrons. These effects result in each excitation corresponding to a different energy, in contrast to Zeeman splitting. These differences could be used to separately address excitations/electrons without the need for additional gates, greatly simplifying the problem of realising a multi-qubit quantum computer. Indeed, each radiation-induced resonance could potentially be considered as a qubit, characterised by the resonant feature shape, frequency and width. The quantum system initial state would be defined by the set(s) of occupied donor sites that minimises the free energy in the system. Manipulation and gate operation could be based on well established NMR protocols.
\begin{acknowledgments}
AR acknowledges financial support from EPSRC and NPL.
DGH would like to acknowledge useful discussions with D.E. Kmelnitskii. The authors would like to thank Stephen Giblin for helpful discussions and careful reading of the manuscript. 
\end{acknowledgments}


\end{document}